\documentclass[aps,prc,reprint,showpacs]{revtex4-1}

\usepackage{amsmath}
\usepackage{bm}
\usepackage{graphicx}
\usepackage{hyperref}
\usepackage{subfigure}
\usepackage{float}
\usepackage{url}

\begin{document}


\title{Boltzmann-Langevin Approach to Pre-equilibrium Correlations in Nuclear Collisions}

\author{Sean Gavin,$^1$ George Moschelli,$^2$ and Christopher Zin$^1$} 
\address{
$^1$Department of Physics and Astronomy, Wayne State University, 
Detroit, MI, 48202\\
$^2$Lawrence Technological University, 21000 West Ten Mile Road, Southfield, MI  48075}

\date{\today}

\begin{abstract}
Correlations born before the onset of hydrodynamic flow can leave observable traces on the final state particles. Measurement of these correlations can yield important information on the isotropization and thermalization process. Starting from a Boltzmann-like kinetic theory in the presence of dynamic Langevin noise, we derive a new partial differential equation for the two-particle correlation function that respects the microscopic conservation laws. We illustrate how this equation can be used to study the effect of thermalization on long range correlations.  
\end{abstract}



\maketitle

%
\section{\label{sec:intro} Introduction}

High energy kinematics and QCD dynamics create correlations between the first partons produced at the beginning of a nuclear collision. Scattering among these partons leads to dissipation that works to erase these correlations, making the system as thermal and locally isotropic as possible. The rapid expansion and short lifetime of the system fight the forces of isotropization, preventing certain correlations from being completely thermalized. Identifying such {\em partially} thermalized correlations can reveal important information about the spacetime character of the thermalization process.  

In this paper we combine the Boltzmann equation in the relaxation time approximation with dynamic Langevin fluctuations to study the effect of thermalization on two-particle correlations. The Boltzmann equation is one of the few tools available  for studying nonequilibrium aspects of ion collisions \cite{Baym:1984np,Banerjee:1989by,Gavin:1990up,Heiselberg:1995sh,Wong:1996va,Nayak:1996ex,Gyulassy:1997ib,Nayak:1997kp,Dumitru:2001kz,Xu:2004mz,Xu:2007aa,Florkowski:2016qig,Hatta:2015kia,Heinz:2015gka,Nopoush:2015yga}.  Nevertheless, the standard form of this equation says nothing about correlations, because of the molecular chaos assumption employed in its description of scattering; see, e.g., \cite{kardar2007statistical}. To describe correlations, we introduce a Langevin noise consistent with the conservation laws obeyed by the microscopic scattering processes \cite{foxUhlenbeck1970A,foxUhlenbeck1970B,bixonZwanzig1969}. We derive a new relativistic transport equation for the two-body distribution function. 

Our interest is driven in part by the discovery of flow-like azimuthal correlations in pA and high-multiplicity pp collisions \cite{Khachatryan:2010gv,Chatrchyan:2013nka,Aad:2013fja,Abelev:2012ola,Adare:2013piz}. Measurements of azimuthal anisotropy in heavy ion collisions provide comprehensive evidence for the hydrodynamic description of these large systems \cite{Shen:2015msa}. The measurement of similar anisotropy in the smaller pp and pA systems raises profound questions about the onset of collective flow and its relation to hydrodynamics.  As a first illustrative application, we study transverse momentum fluctuations, long argued to be a probe of thermalization \cite{Gavin:2003cb}.  These fluctuations have been measured by LHC, RHIC, and SPS experiments -- see Refs.\ \cite{Adams:2005ka,Adams:2005aw,Adams:2004pa,Adamova:2008sx,Abelev:2014ckr,Novak2013,Novak:2013zz} -- for a variety of reasons \cite{Mrowczynski:1998vt,Stephanov:1999zu}. Data markedly deviate from equilibrium expectations in peripheral heavy-ion collisions at LHC and RHIC \cite{Gavin:2011gr}. We argue that measurements in pA collisions can demonstrate whether these systems are indeed thermal.  



%
%
The initial phase space distribution of particles differs in each collision event due largely to the variation of the distribution of nucleons in the colliding nuclei.  These fluctuations introduce observable correlations, since particle pairs are more likely to be found near the ``hot spots'' they produce. In particular, color fields produced by the initial nucleon participants result in hot spots extending across the beam direction at early times \cite{Dumitru:2008wn,Gavin:2008ev}.  These fields produce correlated particles over a broad range in rapidity, likely explaining the ridge and other structure observed in correlation measurements \cite{Wenger:2008ts,Alver:2009id,Daugherity:2008su,Adare:2008ae,Adam:2015gda,Chatrchyan:2012wg,Khachatryan:2016ibd,Aad:2012bu,Wang:2013qca}.

Further dynamic fluctuations occur throughout the evolution of each event due to the stochastic nature of particle interactions.  This thermal noise is a consequence of the same microscopic scattering that produces dissipation and local equilibration. While dissipation tends to dampen the effect of the initial hot spots on final-state particles, noise opposes this dampening. 



This paper is organized as follows.  We will treat the thermalization of correlations using a linearized form of the Boltzmann equation in the relaxation time approximation. In Sec.\ \ref{sec:Beq} we briefly introduce the relativistic Boltzmann equation and discuss its formal solution using the method of characteristics. We focus on the consequences of linearization and the relaxation time approximation on the equation and this solution.

To discuss the evolution of the fluctuating system towards a physically consistent local equilibrium state, we must include dynamic Langevin noise, as pointed out in Refs.\ \cite{Gavin:2003cb,Gavin:2016hmv}. A number of authors have studied theoretical and phenomenological aspects of thermal noise in the context of hydrodynamics \cite{Calzetta:1997aj,Kapusta:2011gt,Kumar:2013twa,Young:2014pka,Yan:2015lfa,Nagai:2016wyx, Gavin:2016hmv,Akamatsu:2016llw}.  
A key motivation for our work here and in \cite{Gavin:2016hmv} is to better understand stochastic hydrodynamic equations.  
In Sec.\ \ref{sec:Roadmap} we use a linearized Boltzmann-Langevin equation to obtain an evolution equation for the two-particle phase space correlation function  that respects the conservation laws, Eq.\ (\ref{eq:G12}). Following Ref.\ \cite{Gavin:2016hmv}, we use analytic techniques for working with stochastic differential equations \cite{van2011stochastic,gardiner2004handbook}. 
In Sec.\ \ref{sec:Collisions} we solve this equation for nuclear collisions using the method of characteristics.  Sections \ref{sec:Roadmap} and \ref{sec:Collisions} constitute the primary results of this paper. 

In Sec.\ \ref{sec:Observables}, we briefly turn to the observable consequences of partial thermalization, where we discuss the long range contribution to transverse momentum fluctuations following Refs.\ \cite{Gavin:2003cb,Gavin:2011gr}.  Our exploratory results suggest striking consequences in pA collisions as these systems approach equilibrium.  

We point out that the Boltzmann equation has also been studied using numerical simulations based on the cascade approach; see, e.g., \cite{Cassing:2009vt,Tindall:2016try,Bass:1998ca,Lin:2004en,Buss:2011mx,Nara:2016phs,Weil:2016zrk,Xu:2004mz}. For these codes to correctly describe dynamic correlations, cross sections for all $m \rightleftharpoons n$ body scattering processes must be specified in accord with detailed balance. This is a tall order, and its difficult to test whether a specific implementation of fluctuations is sufficiently accurate to address any given question of physical interest.    
Complicating matters, simulations can require large statistics to describe some fluctuation observables; see, e.g., \cite{Sharma:2011nj}. 
Our approach can be studied analytically and, therefore, complements such simulations. In particular, Eq.\ (\ref{eq:G12}) integrates out the microscopic sources of fluctuations while retaining their effects at the two-body level.

%
%
%
%
%
%
%
%
%
%
%
%
%
%

\section{Linearized Boltzmann Equation}\label{sec:Beq}
We discuss thermalization in terms of a Boltzmann-like kinetic theory, in which the evolution is characterized by a phase space distribution function $f({\mathbf p},{\mathbf x},t)$ that gives the density of partons of momentum $\mathbf p$ and energy $E$ at the  point $(t, {\mathbf x})$.  In the local-rest frame in which the average momentum density vanishes, the evolution of $f({\mathbf p},{\mathbf x},t)$ is described by the kinetic equation
\begin{equation}\label{eq:Boltzmann}
{\partial \over \partial t}f(\mathbf p,\mathbf x,t) +   
\mathbf v_{\mathbf p}\cdot \bm{\nabla} f(\mathbf p,\mathbf x,t) =   
I\{f\},
\end{equation}
where ${\mathbf v}_{\mathbf p} = {\mathbf p}/E$ is the single particle velocity. The left side of (\ref{eq:Boltzmann}) is a total time derivative of $f$ describing the drift of particles at constant ${\mathbf v}_{\mathbf p}$ between collisions. 

Collisions drive $f$ to the local thermal equilibrium form $f^e$. The corresponding rate of change of $f$ is described by the collision term $I\{f\}$.  For elastic scattering of a single parton species,
\begin{equation}\label{eq:Collisions}
I\{f\} = \int W_{12\rightarrow 34} (f_3f_4-f_1f_2) d{\mathbf p}_2d{\mathbf p}_3d{\mathbf p}_4,
\end{equation}
where $f_i = f({\mathbf p}_i,{\mathbf x},t)$,  $d{\mathbf p} = d^3 p/(2\pi)^3$, and the scattering rate $W_{12\rightarrow 34}\propto \delta(p_1^\mu + p_2^\mu -p_3^\mu - p_4^\mu)$.  Note that (\ref{eq:Collisions}) depends on the products $f_1f_2$ and $f_3f_4$ in accord with the molecular chaos ansatz. A more rigorous description of correlations would replace these products with two-particle distributions.      

Microscopic energy and momentum conservation imply that the moments of $I\{f\}$ with respect to $E$ and  ${\mathbf p}$ must vanish. Elastic scattering also conserves particle number, further requiring that the momentum integral of $I\{f\}$ vanish.  Together, these conservation conditions are
\begin{equation}\label{eq:Constraint}
\int \!d{\mathbf p}\, \left\{ \begin{matrix}1\\{\mathbf p}\\ E\end{matrix} \right\}I\{f\} = 0.
\end{equation}
Furthermore, the structure of (\ref{eq:Collisions}) dictates the momentum dependence of the local equilibrium distribution
\begin{equation}\label{eq:leq}
f^e = e^{-\gamma(E - {\mathbf p}\cdot {\mathbf v} - \mu)/T},
\end{equation}
where the temperature $T$, chemical potential $\mu$ and fluid velocity ${\mathbf v}$ vary in space and time, 
$\gamma = (1-v^2)^{-1/2}$, and we assume Boltzmann statistics.    

In the relaxation time approximation we estimate the collision term (\ref{eq:Collisions}) as
\begin{equation}\label{eq:RTA}
I\{f\}\approx
-{\nu}\left(f(\mathbf p,\mathbf x,t) - f^e(\mathbf p,\mathbf x,t)\right),
\end{equation}
where the relaxation time $\nu^{-1}$ is determined by the microscopic scattering processes.  To be consistent with the conservation conditions (\ref{eq:Constraint}), we require that 
\begin{equation}\label{eq:RTAConstraint}
\int \!d{\mathbf p}\, \left\{ \begin{matrix}1\\{\mathbf p}\\ E\end{matrix} \right\}f  = 
\int \!d{\mathbf p}\, \left\{ \begin{matrix}1\\{\mathbf p}\\ E\end{matrix} \right\}f^e.
\end{equation}
This condition constrains the values of $T$, $\mu$ and ${\mathbf v}$ at each space-time point $({\mathbf x},t)$. 

The relaxation time $\nu^{-1}$ corresponds to the mean free time between parton collisions in a frame where the fluid is locally at rest.  More generally, we write the covariant form of the Boltzmann equation 
\begin{equation}\label{eq:covariantBEQ}
p^\mu\partial_\mu f = -\nu p\cdot u (f-f^e),   
\end{equation}
where the fluid four-velocity is $u^{\mu} = \gamma(1, \mathbf{v})$ and $p\cdot u \equiv p_\mu u^\mu$ for the metric $g^{\mu\nu}={\rm diag}(1,-1,-1,-1)$. This equation reduces to (\ref{eq:Boltzmann}) with (\ref{eq:RTA}) in the local rest frame where $u_\mu = (1,0,0,0)$. 

To simplify this equation we introduce a proper time parameter $\tau$ defined by the differential equation
\begin{equation}\label{eq:path}
\frac{dx^\mu}{d\tau} = \frac{p^\mu}{p\cdot u}. 
\end{equation}
The time component of (\ref{eq:path}), $dt/d\tau = E/p\cdot u$, implies that $\tau$ is the time in the fluid rest frame. Moreover, (\ref{eq:path}) defines the path $x^\mu(\tau)$ of the center of momentum of a phase space cell of mean $p^\mu$ relative to this frame. 
We now write the Boltzmann equation as
\begin{equation}\label{eq:pathBEQ}
df/d\tau = -\nu (f-f^e).
\end{equation}
This reduction of a first order partial differential equation to a set of ordinary differential equations is a classic application of the method of characteristics \cite{stone2009mathematics}. This method is often used to solve the nonrelativistic Boltzmann equation \cite{reif2009fundamentals}.  

In solving (\ref{eq:pathBEQ}), we start by considering the free streaming case in which there are no collisions, i.e., we take the right side of (\ref{eq:pathBEQ}) to be zero. Equation (\ref{eq:path}) implies that the matter in a cell initially at $\mathbf{x}_0$ drifts unchanged along the trajectory $\mathbf{x} = \mathbf{x}_0 + \mathbf{v}_{\mathbf{p}} t$. We use (\ref{eq:pathBEQ}) to find
\begin{eqnarray}\label{eq:freeStreaming}
f({\mathbf p},{\mathbf x},\tau) &=& f_0({\mathbf p},{\mathbf x} - \mathbf{v}_{\mathbf{p}} t), \,\,\,\,{\rm free\ streaming}
\end{eqnarray}
where $t$ is determined as a function of $\tau$ by $dt/d\tau = E/p\cdot u$ and $f_0({\mathbf p},{\mathbf x})$ is the initial distribution. It is clear that (\ref{eq:freeStreaming}) is a solution of (\ref{eq:Boltzmann}). 

In the presence of both collisions and drift, we write (\ref{eq:pathBEQ}) as the integral equation  
\begin{eqnarray}\label{eq:pathIntegral}
f &=& f_0({\mathbf p},{\mathbf x} - \mathbf{v}_{\mathbf{p}} t)S(\tau,\tau_0)\nonumber\\ 
&+& \int_{\tau_0}^{\tau} d\tau^\prime \nu(\tau^\prime) S(\tau,\tau^\prime)
f^e({\mathbf p},{\mathbf x} - \mathbf{v}_{\mathbf{p}} t^\prime), 
\end{eqnarray}
where $t = t(\tau)$ and $t^\prime = t(\tau^\prime)$ satisfy (\ref{eq:path}). We define the survival probability as
\begin{equation}\label{eq:survival}
S(\tau, \tau_0) = \exp\{-\int_{\tau_0}^{\tau}  \nu(\tau^\prime)d\tau^\prime\},
\end{equation}
the probability that partons suffer no collisions as they travel along their characteristic path. To compute (\ref{eq:pathIntegral}) we must specify the parameters $T$, $\mathbf v$ and $\mu$ as a function of time by enforcing the nonlinear constraint  (\ref{eq:RTAConstraint}).  

Baym solved the Boltzmann equation in the relaxation time approximation (\ref{eq:covariantBEQ}) assuming longitudinal boost-invariant expansion and neglecting transverse flow \cite{Baym:1984np}.  In our formulation, these additional assumptions and (\ref{eq:path}) imply that motion along the $z$ direction starting at $z_0$ satisfies $p_z(t-t_0) = E(z-z_0)$ for $\tau= (t^2-z^2)^{1/2}$.  Boost invariance along $z$ further restricts $f$ to be a function only of $p_z^\prime = (tp_z-zE)/\tau$, the transverse momentum $p_t$, and $\tau$. It follows that $p_z\tau = p_{z0}\tau_0$.  We see that the free streaming case gives  $f = f_0(p_z\tau/\tau_0, \tau)$, while (\ref{eq:pathIntegral})  more generally gives Baym's eq.\ (17).  
    
In this paper we will also use the linearized versions of these equations. We expand $f \approx f^e(1+h)$, where $f^e$ is given by (\ref{eq:leq}) and $h\ll 1$ is a small perturbation. 
We linearize the collision term 
\begin{equation}\label{eq:LinearCollisions}
I\{f\} \approx \int_{2,3,4} W_{12\rightarrow 34} f_1^e f_2^e (h_3+h_4-h_1-h_2)  \equiv Lh,
\end{equation}
where the integrations are again over momentum.  Consider the eigenfunctions of this operator, which satisfy $L\phi_\alpha = -\nu_\alpha \phi_\alpha$.  The first five eigenfunctions have the eigenvalue zero and are linear in the conserved quantities $1$, ${\mathbf p}$ and $E$. The linear combinations
\begin{equation}\label{eq:eigenfunctions}
\phi_1 = 1, \,\,\,\, \phi_{2,3,4} = \sqrt{\frac{n}{wT}}{\mathbf p},  \,\,\,\,\phi_{5} = \sqrt{\frac{n}{c_vT}}\left(E-\frac{e}{n}\right) 
\end{equation}
are orthonormal in the sense that 
\begin{equation}\label{eq:orthogonality}
\int \!d{\mathbf p}\, f^e \phi_\alpha \phi_\beta = n \delta_{\alpha\beta}.  
\end{equation}
The other eigenvalues are positive.  

For the linearized form $f\approx f^e(1+h)$, the conservation conditions for the relaxation time approximation (\ref{eq:RTAConstraint}) become
\begin{equation}\label{eq:LRTAConstraint}
\int \!d{\mathbf p}\, \phi_\alpha f^e h  = 0 \,\,\,\,\,\,\,\,\,\,{\rm for}\,\,\,\,\,\,\,\,\,\,\alpha = 1,\ldots 5.
\end{equation}
i.e., the first five eigenfunctions are orthogonal to the perturbation $h$.  We see that the linearized condition (\ref{eq:LRTAConstraint}) does not specify the values of $T$, $\mathbf v$, and $\mu$ as with the exact condition (\ref{eq:Constraint}), but in contrast with (\ref{eq:RTAConstraint}).   

To specify the local equilibrium parameters in (\ref{eq:leq}), we require that $f^e$ satisfy the Boltzmann equation with $I\{f\}= 0$, so that
\begin{equation}\label{eq:pathBEQe}
df^e/d\tau = 0, 
\end{equation}
where the parameters $T$, $\mathbf v$ and $\mu$ depend on position along the path $x^\mu(\tau)$ from (\ref{eq:path}). The evolution of $f^e$ describes the flow of a dissipation-free fluid, as we now demonstrate.  Multiplying (\ref{eq:Boltzmann}) by $p^\nu$, integrating over momentum, and enforcing the condition (\ref{eq:Constraint}) gives $\partial_\mu T^{\mu\nu} = 0$, where  $T^{\mu\nu} = \int d{\mathbf p} f p^\mu p^{\nu}/E$ is the stress energy tensor. Integrating (\ref{eq:Boltzmann}) without a factor and enforcing (\ref{eq:Constraint}) yields $\partial_\mu j^{\mu} = 0$, where  $j^{\mu} = \int d{\mathbf p} f p^\mu /E= n u^\mu$ is the parton current for parton density $n$.  When $f =f^e$ given by (\ref{eq:leq}), we obtain the stress-energy tensor for an ideal dissipation-free fluid, $T_{\rm id}^{\mu\nu} = (e+p)u^\mu u^\nu - p g^{\mu\nu}$, where $e$ is the energy density and $p$ is the pressure for an ideal Boltzmann gas. The equations of motion for this system are therefore equivalent to relativistic Euler equations.  Observe that the equation for the full distribution $f$  includes dissipation at linear order.  


The relaxation time approximation amounts to the assumption that the eigenfunctions of $L$ with $\nu_\alpha \ne 0$ have a common value $\nu$.  To explicitly enforce the condition (\ref{eq:LRTAConstraint}), we write (\ref{eq:RTA}) as
\begin{equation}\label{eq:LRTA}
I\{f\}\approx  
-{\nu}(1-P) f(\mathbf p,\mathbf x,t),
\end{equation}
where $P$ is a projection operator that projects $f$ into the corresponding local equilibrium distribution $f^e$.  We define
\begin{equation}\label{eq:Projector}
P\psi(\mathbf{p})  = \frac{f^e(\mathbf{p})}{n} \sum_{\alpha = 1}^5 \phi_\alpha(\mathbf{p}) \int \!d{\mathbf p}^\prime \, \phi_\alpha(\mathbf{p}^\prime)\psi(\mathbf{p}^\prime),
\end{equation}
where $\psi$ is an arbitrary function of momentum. As a projection operator, $P$ satisfies $P^2 = P$ as well as $P(1-P) = 0$.  We use (\ref{eq:orthogonality}) and (\ref{eq:LRTAConstraint}) together with the explicit eigenfunctions (\ref{eq:eigenfunctions}) to show that $Pf = f^e$, with corrections beyond linear order.   

The linearized Boltzmann equation is then  
\begin{equation}\label{eq:LRTAequation}
df/d\tau = -{\nu}(1-P) f. 
\end{equation}
Observe that $P$ commutes with $d/d\tau$ because of (\ref{eq:pathBEQe}) and (\ref{eq:Projector}). Multiplying both sides of (\ref{eq:LRTAequation}) by $1-P$ and using $(1-P)^2 = 1-P$ gives $d(1-P)f/d\tau = -\nu(1-P)f$, which has the solution $(1-P)f = (1-P)f_0 S(\tau, \tau_0)$. On the other hand, multiplying  (\ref{eq:LRTAequation}) by  $P$ implies $dPf/d\tau = 0$.  Identifying the constant $Pf$ as the local equilibrium distribution $f^e$, we see that this equation is equivalent to (\ref{eq:pathBEQe}). We identify $(1-P)f$ as the deviation from local equilibrium $f -f^e$.

We find
%
%
\begin{eqnarray}\label{eq:pathIntegralLinear}
f &=& f_0({\mathbf p},{\mathbf x} - \mathbf{v}_{\mathbf{p}} t)S(\tau,\tau_0)\nonumber\\ 
&+& 
f^e({\mathbf p},{\mathbf x} - \mathbf{v}_{\mathbf{p}} t)(1-S(\tau,\tau_0)),
\end{eqnarray}
where $t$ is a function of $\tau$ in accord with  (\ref{eq:path}). 
As a check, we can obtain this result without using the $P$ operator by noting that (\ref{eq:pathBEQe}) implies that the linearized $f^e$ is constant in $\tau$, so that we can integrate (\ref{eq:pathIntegral}) by parts.  

The factor $S$ that determines the extent of thermalization (\ref{eq:survival}) is the same as in the fully nonlinear relaxation time approach.  This factor is only a function of the proper time $\tau$ and the collision frequency $\nu$.  

We use the linearized relaxation time approximation in the next section because it provides a simple description of transport that incorporates the conservation laws effectively. While it might not describe the first instants of pre-equilibrium evolution as effectively as the full relaxation time approach or the full Boltzmann equation, none of these approaches is fully reliable at that stage. 


\section{Dynamic Fluctuations}\label{sec:Roadmap}
In concert with the relaxation process described by (\ref{eq:Collisions}) and (\ref{eq:RTA}), scattering causes stochastic fluctuations of the phase space distribution. These fluctuations give rise to correlations in addition to those already present in the initial conditions.  We characterize these correlations using 
\begin{eqnarray}
C_{12}\equiv C(\mathbf{p}_1,\mathbf{x}_1,\mathbf{p}_2,\mathbf{x}_2,t) 
 = \langle f_1f_2\rangle - \langle f_1\rangle\langle  f_2\rangle,
\label{eq:CcorrF2}
\end{eqnarray}
%
where $f_i= f(\mathbf{p}_i,\mathbf{x}_i,t)$ 
and the brackets denote an average over an ensemble of possible fluctuations with fixed initial conditions. 
If we were to omit thermal noise, $C_{12}$ would vanish in equilibrium. We refer to the average in (\ref{eq:CcorrF2}) as the ``noise average'' or the ``thermal average.''

To incorporate fluctuations into our kinetic theory approach, we build on the theory of Brownian motion \cite{van2011stochastic,gardiner2004handbook}. One describes the erratic fluctuations of a single Brownian particle suspended in a fluid using a Langevin equation $m\dot{v} = - m\gamma v  + F$. Microscopic collisions with atoms in the fluid create a stochastic force $F$ that generates each jump, together with the friction coefficient $\gamma$ that dissipates the subsequent motion. We write this equation as a difference equation 
\begin{equation}\label{eq:Brownian1}
v(t+\Delta t)-v(t) \equiv \Delta v =  - \gamma v(t)\Delta t  +  \Delta W, 
\end{equation}
where  $\Delta W$ represents the net change in $v$ due to collisions in the time interval from $t$ to $t+\Delta t$.  Each collision is independent of the others in direction and magnitude, so that 
\begin{equation}\label{eq:Brownian2}
\langle \Delta W\rangle  = 0\,\,\,\,\,\,\,\,\,\,\,\,  {\rm and} \,\,\,\,\,\,\,\,\,\,\,\, \langle \Delta W^2\rangle  = \Gamma \Delta t, 
\end{equation}
when averaged over the noise, i.e., all possible trajectories of the Brownian particle starting with the same velocity $v$ and position $x$. The linear relation $\langle \Delta W^2\rangle  \propto \Delta t$ is typical of random-walk processes and, unopposed by friction, would cause the variance of $v$ to increase in proportion to time \cite{gardiner2004handbook}. In Brownian motion the fluctuation-dissipation theorem fixes the coefficient $\Gamma  =  2\gamma \langle v^2\rangle =2\gamma T/m$  by requiring that fluctuations in equilibrium have the appropriate thermodynamic limit.  

To add Langevin noise to the linearized Boltzmann equation, we divide phase space into discrete cells.  The phase space population fluctuates due to the action of collisions, which randomly transfer momentum between particles in cells centered at the same point in space. To describe this process we write (\ref{eq:LRTAequation}) as a difference equation
\begin{equation}\label{eq:LRTAequation1}
f(\tau+\Delta \tau) - f(\tau)\equiv \Delta f = -{\nu}(1-P) f(\tau) d\tau + \Delta W,
\end{equation}
where $\Delta W$ represents the stochastic increment to the phase space density $f$ at $(\mathbf{p}, \mathbf{x})$  from $\tau$ to $\tau+\Delta \tau$. These increments satisfy 
\begin{equation}\label{eq:BLnoise}
\langle \Delta W(p_1, x_1) \Delta W(p_2, x_2) \rangle = \Gamma_{12}\Delta\tau;
\end{equation}
see, e.g., \cite{gardiner2004handbook}.

To obtain the differential equation for the linearized one-body distribution $\langle f(\tau)\rangle$, we average (\ref{eq:LRTAequation1}) over the noise to find $\langle f(\tau+\Delta \tau)\rangle - \langle  f(\tau)\rangle = -\nu(1-P) \langle f(\tau)\rangle \Delta \tau$, so that
\begin{equation}\label{eq:BLavg}
d{\langle f\rangle}/d\tau = - \nu(1-P) \langle f\rangle
\end{equation}
as $\Delta \tau \rightarrow 0$. In the long time limit, the average $\langle f\rangle$ follows the solution (\ref{eq:pathIntegralLinear}). The noise term has no effect on the mean. Observe that  we will later consider the more general possibility that $\langle f\rangle$ satisfies the nonlinear equation (\ref{eq:pathIntegral}).

We stress that the linearized $f$ in each event in this average has the same initial conditions and, therefore, the same local equilibrium state $Pf = f^e$. The linearized evolution of $f^e$ follows from the Euler equation, which omits both dissipation and fluctuations. Similarly, drift follows the deterministic paths $x^\mu(\tau)$ described by (\ref{eq:path}) for fixed initial conditions. Both $f^e$ and the paths would differ from event to event.      

We now construct a differential equation for the correlation function (\ref{eq:CcorrF2}) following the procedure of Ref.\ \cite{Gavin:2016hmv}. We take the product of (\ref{eq:LRTAequation}) at two phase space points  $\Delta (f_1f_2) = f_2\Delta f_1 + f_1\Delta f_2 + \Delta f_1 \Delta f_2$. The average of the first two terms is $-[\nu(1-P_1)+\nu(1-P_2)]\langle f_1f_2\rangle \Delta\tau$.  Averaging the third term using (\ref{eq:BLnoise}),  we find $\langle \Delta f_1 \Delta f_2\rangle = \langle \Delta W_1 \Delta W_2\rangle = \Gamma_{12}\Delta \tau$ to leading order in $\Delta \tau$. We combine these contributions and take  $\Delta \tau \rightarrow 0$ to obtain 
\begin{equation}\label{eq:BLsquared}
{{d}\over {d\tau}}\langle f_1f_2\rangle  =  -[\nu(1-P_1)+\nu(1-P_2)]\langle f_1f_2\rangle  +  \Gamma_{12}.  
\end{equation}
In the theory of stochastic differential equations, the need to include $\Delta f_1 \Delta f_2$ in $\Delta (f_1f_2)$ when noise is present is known as the It$\hat{\rm{o}}$ product rule. We combine  (\ref{eq:BLavg}) and (\ref{eq:BLsquared}) to write
\begin{equation}\label{eq:BLC}
\left({{d}\over {d\tau}}  + \nu(1-P_1)+\nu(1-P_2)\right) C_{12} = \Gamma_{12},  
\end{equation}
where $C_{12} =\langle f_1f_2\rangle -\langle f_1\rangle \langle f_2\rangle$ is the correlation function (\ref{eq:CcorrF2}).  

To understand the observable impact of correlations, we turn to the related correlation function 
\begin{eqnarray}
G_{12} = C_{12} - \langle f_1\rangle\delta(1-2),
\label{eq:Gcorr}
\end{eqnarray}
where we abbreviate $\delta(1 - 2)\equiv \delta(\mathbf{x}_1 - \mathbf{x}_2)\delta(\mathbf{p}_1 - \mathbf{p}_2)$. 
The quantity $G_{12}$ compares the phase space density of distinct pairs, $\langle f_1 f_2\rangle - \langle f_1\rangle \delta(1- 2)$, to the expectation $\langle f_1\rangle \langle f_2\rangle$ in the absence of correlations. In principle, one can measure $G_{12}$ simply by counting pairs of particles. We use (\ref{eq:BLC}) to find
\begin{equation}\label{eq:BLG}
\left({{d}\over {d\tau}}  + \nu(1-P_1)+\nu(1-P_2)\right) G_{12} = \Gamma_{12}^\prime,  
\end{equation}
where 
\begin{equation}\label{eq:BLGnoise}
\Gamma_{12} - \Gamma_{12}^\prime = \left({{d}\over {d\tau}} +  2\nu(1-P_1)\right)\langle f_1\rangle\delta(1-2).     
\end{equation}
We derived analogous equations for two-particle correlation functions in the hydrodynamic regime in Ref.\ \cite{Gavin:2016hmv}. Up to this point, the derivations have been quite similar.   

The pair correlation function $G_{12}$ vanishes in local equilibrium in a sufficiently large system. Particle number fluctuations locally satisfy Poisson statistics in equilibrium if the  grand canonical ensemble applies.   Number fluctuations then satisfy $\langle N^2\rangle - \langle N\rangle^2 = \langle N\rangle$, so that the equilibrium phase space correlations satisfy $(\langle f_1f_2\rangle -\langle f_1\rangle \langle f_2\rangle)^e = \langle f_1\rangle \delta(1 - 2)$. 

We now turn to obtain the coefficient $\Gamma_{12}$.  We can infer many features of $\Gamma_{12}$ from first principles. First, the stochastic nature of fluctuations implies that $\Delta W_1$ and $\Delta W_2$ are uncorrelated for different cells  $(\mathbf{p}_1, \mathbf{x}_1)$ and $(\mathbf{p}_2, \mathbf{x}_2)$. We therefore expect $\Gamma_{12}$ to be singular at $(\mathbf{p}_1, \mathbf{x}_1) = (\mathbf{p}_2, \mathbf{x}_2)$ as the cell size tends to zero, and zero otherwise \cite{gardiner2004handbook}. Second, we expect $\Gamma_{12}$ to vanish in local equilibrium, since correlations come from scattering and detailed balance implies $\left(\partial f/ \partial t\right)_{\rm coll}\equiv 0$. We therefore expect
\begin{equation}\label{eq:noiseExpect}
\Gamma_{12} = (1-P_1)(1-P_2)a_1\delta(1-2),
\end{equation}
where $a_1$ is a function to be determined. 
Combining (\ref{eq:BLC}) and (\ref{eq:noiseExpect}) and multiplying by $P_1P_2$, we find 
\begin{equation}\label{eq:ConstantPairs}
{{d}\over {d\tau}} P_1P_2 C_{12} = 0,  
\end{equation}
where we used the property $P(1-P) = 0$ of projection operators and the fact that $P$ commutes with $d/d\tau$ due to (\ref{eq:pathBEQe}) and (\ref{eq:Projector}).   

We use the fluctuation-dissipation theorem to determine the coefficient $\Gamma_{12}$ near equilibrium. 
To begin, consider a uniform system. In equilibrium, the derivative in (\ref{eq:BLC}) must vanish, so that 
\begin{equation}\label{eq:Brownian4}
\Gamma_{12}  =  [\nu(1-P_1)+\nu(1-P_2)] C_{12}^e.
\end{equation}
We then write
\begin{equation}\label{eq:FDTuniform}
\Gamma_{12}^e =  2\nu(1-P_1)(1-P_2)\langle f_1\rangle\delta(1 - 2),
\end{equation}
where we use $1-P_1 =(1-P_1)^2$ and exploit the delta function.  In a uniform system where $\langle f_1\rangle = f^e$, this quantity is strictly zero, although it has the correct general structure.

Now consider the steady state behavior of a system that {\em cannot} equilibrate due to large gradients maintained, e.g., by fixed boundary conditions. In this case the $\tau$ derivatives do not vanish due to the $v_1\cdot\nabla_1 + v_2\cdot\nabla_2$  contributions. With such large gradients we  must use  (\ref{eq:pathBEQ}) to describe $\langle f\rangle$ instead of the linearized Eq.\  (\ref{eq:BLavg}). In this case $Pf \neq f^e$.  
We operate on (\ref{eq:BLG}) with $P_1P_2$ and use (\ref{eq:Gcorr}) and (\ref{eq:ConstantPairs}) to obtain   
\begin{eqnarray}\label{eq:GNoiseTrick}
 P_1P_2\Gamma_{12}^\prime &=& -P_1P_2{{d}\over {d\tau}} \langle f_1\rangle\delta(1 - 2)\nonumber\\ 
 &=& \nu P_1P_2(\langle f_1\rangle - f_1^e)\delta(1 - 2).  
\end{eqnarray}
In the last step we used the full (\ref{eq:pathBEQ}) to evaluate the derivative, because the constrained system is never close to $f^e$. We use (\ref{eq:BLGnoise}) with (\ref{eq:pathBEQ}) to obtain 
\begin{equation}\label{eq:Cnoise}
\Gamma_{12} = (1-P_1)(1-P_2)(\langle f_1\rangle + f_1^e)\delta(1-2),
\end{equation}
which reduces to the uniform value (\ref{eq:FDTuniform}) when the boundary constraints are removed.  
We find 
\begin{equation}\label{eq:GNoiseF}
\Gamma_{12}^\prime = \nu P_1P_2(\langle f_1\rangle - f_1^e) \delta(1 - 2)
\end{equation}
by applying $(1-P_1)(1-P_2)$ to (\ref{eq:BLGnoise}). 

We now write the general evolution equation for the two-body correlation function 
\begin{eqnarray}\label{eq:G12}
\left({{d}\over {d\tau}}  +\nu(1-P_1)+\nu(1-P_2)\right) G_{12}\nonumber\\ 
= \nu P_1P_2(\langle f_1\rangle - f_1^e)\delta(1-2), 
\end{eqnarray}
where the presence of projection operators enforces energy, momentum and number conservation. For completeness, we explicitly exhibit the drift terms in a local rest frame and write 
\begin{eqnarray}\label{eq:G12all}
\left(\frac{\partial}{\partial t} + {\mathbf{v}}_{\mathbf{p}_1}\cdot {\bm{\nabla}}_1 + {\mathbf{v}}_{\mathbf{p}_2}\cdot {{\bm\nabla}}_2 + \nu(2 - P_1 -P_2) \right) G_{12}\nonumber\\ 
= \nu P_1P_2(\langle f_1\rangle - f_1^e)\delta(1-2),\quad\quad 
\end{eqnarray}
where the relaxation rate $\nu$ and projection operators $P_i$ depend on the average one-body distribution $\langle f(\mathbf{p},\mathbf{x},t) \rangle$ and the local equilibrium distribution $f^e$. Equation (\ref{eq:G12all}) was derived earlier by Dufty, Lee and Brey in Ref.\ \cite{PhysRevE.51.297} for non-relativistic fluids from a general analysis of the BBGKY hierarchy.  

How might we use these equations in phenomenological applications? To solve (\ref{eq:G12}), we start with the initial condition corresponding to a single collision event. We can then use (\ref{eq:pathIntegral}) to solve the one body Boltzmann equation for $\langle f(\mathbf{p},\mathbf{x},t) \rangle$ together with the conservation conditions (\ref{eq:RTAConstraint}) that fix the parameters $T, \mathbf{v}$ and $\mu$ in the local equilibrium distribution $f^e$. We then solve (\ref{eq:G12}) for the correlation function. 
One must then average over an ensemble of initial conditions.  Physically, the difference between $\langle f\rangle$ and $f^e$ may be arbitrarily large; only the fluctuations $f-\langle f\rangle$ need be small.  In fact, such general solutions need not ever reach equilibrium as discussed, e.g., in Ref.\ \cite{Gavin:1990up}.  

%

For the illustrative calculations in Sec.\ \ref{sec:Observables}, we assume that the departures from local equilibrium are always small enough that the linearized solution  (\ref{eq:pathIntegralLinear}) for $\langle f\rangle$ is applicable. In this case, one can solve dissipation-free Euler equations to determine effective $T, \mathbf{v}$ and $\mu$ parameters for the initial conditions in each event (although we will not need to do that explicitly here).  In this case, the source term in (\ref{eq:G12}) vanishes identically.  

An early effort to study the Boltzmann-Langevin equation in a relativistic context is Ref.\ \cite{Calzetta:1999xh}. Our early effort to address the thermalization using this equations outlined the path followed here \cite{Gavin:2003cb}. The effects of critical phenomena were introduced in Ref.\ \cite{Stephanov:2009ra}, but spatial inhomogeneity was not considered.

\section{Ion Collisions}\label{sec:Collisions}
We are interested in observable consequences of pre-equilibrium correlations that depend on the correlation function $G_{12}$, 
\begin{eqnarray}
G_{12} = \langle f_1f_2\rangle -\langle f_1\rangle\langle f_2\rangle - \langle f_1\rangle\delta(1-2).
\label{eq:Gcorr2}
\end{eqnarray}
In this section we construct formal solutions for the evolution of $G_{12}$.  In Sec.\  \ref{sec:Observables} we use this solution for an approximate analysis of transverse momentum fluctuations as an example of the kind of phenomenological questions one might address.  

We solve (\ref{eq:G12}) following our derivation of (\ref{eq:pathIntegralLinear}), by multiplying (\ref{eq:G12}) by the four combinations of the projection operators $P_i$ and $1-P_j$.  Operating on (\ref{eq:G12}) with $(1-P_1)(1-P_2)$ gives
\begin{eqnarray}\label{eq:deltaG1200}
{{d}\over {d\tau}} [(1-P_1)(1-P_2)G_{12}] \quad\quad\quad \quad\quad \quad\quad \quad    \nonumber\\= -2\nu (1-P_1)(1-P_2)G_{12}.    
\end{eqnarray}
If $\delta f =f-f^e$ is the deviation of the phase space distribution from its local equilibrium value, then one can interpret $(1-P_1)(1-P_2)G_{12}$ as the correlation function $\langle \delta f_1\delta f_2\rangle - \langle \delta f_1\rangle\langle\delta f_2\rangle- \langle\delta f_1\rangle\delta(1-2)$.
This non-equilibrium contribution to correlations then satisfies
\begin{equation}\label{eq:deltaG12}
\frac{d}{d\tau }\Delta G_{12} = - 2\nu \Delta G_{12}.    
\end{equation}
Repeating this process with the other operator combinations,  we define the functions:  
\begin{eqnarray}\label{eq:DeltaGdef}
         G^e_{12} = P_1P_2G_{12},\quad \Delta G_{12} = (1-P_1)(1-P_2)G_{12}\quad
 \end{eqnarray}
and 
\begin{eqnarray}\label{eq:Xdef}
X_{12} =   (1-P_1)P_2G_{12}. 
\end{eqnarray}
The observable correlation function is the sum of these functions 
\begin{eqnarray}\label{eq:Gtotal}
         G_{12}= G^e_{12} + X_{12} + X_{21} + \Delta G_{12} 
 \end{eqnarray}
because of the identity
$1 = P_1P_2+P_1(1-P_2) + (1-P_1)P_2 +(1-P_1)(1-P_2)$. 

The mixed correlation function $X_{12}$ is the covariance $\langle \delta f_1f^e_2\rangle - \langle \delta f_1\rangle f^e_2$.   We find 
\begin{equation}\label{eq:X12}
dX_{12}/d\tau  = - \nu X_{12}. 
\end{equation}
Observe that the noise average $\langle \delta f_1f^e_2\rangle$ need not equal $\langle \delta f_1\rangle f^e_2$, because $f^e$ and $\delta f$ correspond to the same $T$, $v$ and $\mu$. In essence, $X_{12}$ enforces the conservation laws.


To analyze local equilibrium correlations,  we use (\ref{eq:ConstantPairs}) to write 
\begin{equation}\label{eq:GE12}
dC^e_{12}/d\tau = 0,   
\end{equation}
where $C^e_{12} = G^e_{12} + \langle f_1\rangle \delta(1-2)$. If the fully linearized solution (\ref{eq:pathIntegralLinear}) for $\langle f\rangle$ is applicable, then  (\ref{eq:pathBEQe}) holds. In this case $G^e$ is constant along the characteristics as are $C^e$ and $f^e$. We will assume this to be the case in the next section.  However, we point out that a more general non-linear description of the underlying flow described by (\ref{eq:covariantBEQ}) would allow $G^e$ to vary with $\tau$.  
[For completeness, observe that we can extract (\ref{eq:GE12}) from (\ref{eq:G12}) by multiplying by $P_1P_2$.  This gives $dG^e_{12}/d\tau = \nu P_1P_2(\langle f_1\rangle - f_1^e)\delta(1-2)$.  We then use (\ref{eq:pathBEQe}) to identify the right side as $-P_1P_2d\langle f_1\rangle/d\tau \delta(1-2)$ and obtain (\ref{eq:GE12}).]  

We construct solutions following the one body linearized solution (\ref{eq:pathIntegral}) by integrating (\ref{eq:DeltaGdef}), (\ref{eq:X12}) and (\ref{eq:GE12}) and assembling the parts using (\ref{eq:Xdef}). We obtain
\begin{eqnarray}\label{eq:pathIntegral2}
G_{12}= G_{12}^e + (X_{12}^0+X_{21}^0)S + \Delta G_{12}^0S^2
\end{eqnarray}
where the survival probability $S$ is given by (\ref{eq:survival}). The local equilibrium function has arguments
\begin{eqnarray}\label{eq:Gpath}
G_{12}^e= G^e({\mathbf p}_1,{\mathbf x}_{1}-{\mathbf{v}}_{\mathbf{p}_1}t,{\mathbf p}_2,{\mathbf x}_{2}-{\mathbf{v}}_{\mathbf{p}_2}t), 
\end{eqnarray}
where drift for the two particle correlation function again follows (\ref{eq:path}).  The temperature and other local equilibrium parameters in these linearized equations follows the relativistic Euler equation. The initial functions  $X_{12}^0$, and $\Delta G_{12}^0$ follow a similar path dependence. Their values are determined by the initial spatial distribution of nucleon participants and their first few interactions.  

\section{Observing Thermalization}\label{sec:Observables}
Measurements of $p_t$ fluctuations can probe the onset of thermalization.  
In Ref.\ \cite{Gavin:2003cb} we proposed studying thermalization by measuring the centrality dependence of the covariance
 \begin{equation}\label{eq:ptFluctExp}
 \langle \delta p_{t1}\delta p_{t2}\rangle = 
{ {\langle \sum_{i \neq j}\delta p_{t,i}\delta p_{t,j}\rangle}\over{\langle N(N-1)\rangle}},
 \end{equation}
where $i$ and $j$ label a distinct pair of particles  from each event, $\delta p_{ti} = p_{ti}-\langle p_t\rangle$, and the brackets represent the average over events. The average transverse momentum is $\langle p_t\rangle =  \langle P_t \rangle /\langle N\rangle$, where $P_t =  \sum_i p_{t,i}$  the total momentum in an event. The mean number of pairs is $\langle N(N-1)\rangle$. 

\begin{figure}[bt]
\includegraphics[width = \linewidth]{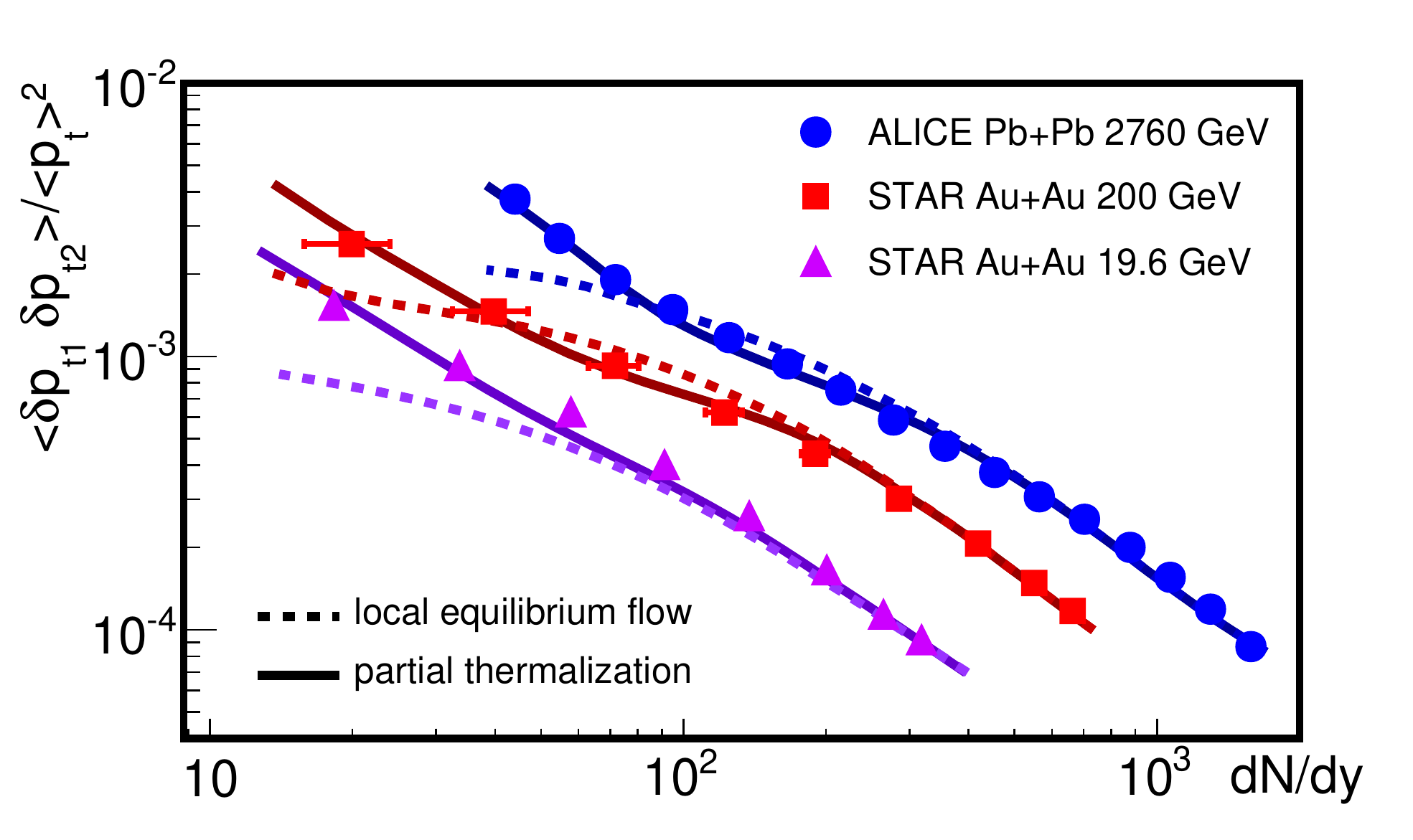}%
\caption{(color online) Transverse momentum fluctuations as a function of the charged-particle rapidity density $dN/dy$ for partial thermalization (solid curves) and local equilibrium flow (dashed curves). 
Data (circles, squares, and triangles) are from Refs.\  \cite{Adams:2005ka}, \cite{Abelev:2014ckr}, and \cite{Novak2013,Novak:2013zz}, respectively.  
\label{fig:Heavy}}
\end{figure}

Experimental results for $\langle \delta p_{t1}\delta p_{t2}\rangle$ are shown in Fig.\ \ref{fig:Heavy} at three beam energies \cite{Adams:2005ka,Abelev:2014ckr,Novak2013,Novak:2013zz}. We indicate the intensity of the nuclear collisions using the rapidity density of charged particles $dN/dy$ to allow eventual comparison to pp or pA collisions \cite{ALICE:2012xs}.  In the heavy Pb+Pb and Au+Au systems, $dN/dy$ is tightly correlated with impact parameter $b$.  We convert $dN/dy$ to the number of participants $N_p(b)$ using data from Refs.\ \cite{Abelev:2008ab,Schukraft:2011kc,Aamodt:2010cz,Alver:2010ck}. We remark that the measured $\langle \delta p_{t1}\delta p_{t2}\rangle/\langle p_t\rangle^2$  for different energies lie on top of each other when plotted as functions of $N_p$. Using $dN/dy$ separates the energies for individual inspection.  %

%
%
%
%

%
%

To compute the effect of thermalization on $p_t$ fluctuations, we will derive the following expression 
\begin{eqnarray}\label{eq:dptEvolution}
  \langle \delta p_{t1}\delta p_{t2}\rangle &=&  \langle \delta p_{t1}\delta p_{t2}\rangle_e \nonumber\\
  &+& \int\!\! \delta p_{t1}\delta p_{t2} \frac{\langle G_{12}- G_{12}^e\rangle}{\langle N(N-1)\rangle}  d\omega_1d\omega_2,\,\,\,\,\,
\end{eqnarray}
%
where $G_{12}$ is the noise-averaged correlation function (\ref{eq:Gcorr}). We abbreviate the differential phase space elements $d\omega = d\mathbf{x}d\mathbf{p}$, where the spatial integrals are on the Cooper-Frye freeze-out surface with $dx=p^\mu d\sigma_\mu/E$ \cite{Cooper:1974mv}. We will use the solution (\ref{eq:dptEvolution}) to compute the deviation of $G_{12}$ from its local equilibrium value $G_{12}^e$ at freeze out. The quantity $\langle \delta p_{t1}\delta p_{t2}\rangle_e$ includes both thermal and initial state fluctuations for a system in local equilibrium.  

In this section we must distinguish the event averages in (\ref{eq:ptFluctExp}) and (\ref{eq:dptEvolution}) from the average over thermal fluctuations in the previous sections. From here on, we denote the thermal noise average as $\langle X\rangle_n$.  The average over events of a noise-averaged quantity $\langle \langle X\rangle_n\rangle$ is equivalent to an average of $\langle X\rangle_n$ over the initial conditions. 

To derive (\ref{eq:dptEvolution}),  we write the unrestricted sum in an event $\sum_{i,j} \delta p_{ti}\delta p_{tj} =  \int \delta p_{t1} \delta p_{t2} \langle f_1f_2\rangle_n d\omega_1d\omega_2$.  We obtain the restricted sum in (\ref{eq:ptFluctExp}) by subtracting  $\sum_{i} (\delta p_{ti})^2 = \int \delta p_{t1} \delta p_{t2}  \langle f_1\rangle_n \delta(1-2) d\omega_1d\omega_2$.  Next, we average this quantity over events. To write the result in terms of $G_{12}$, we add and subtract a $\langle f_1\rangle_n\langle f_2\rangle_n$ term and use (\ref{eq:Gcorr2}) to find
\begin{eqnarray}\label{eq:unres1}
\langle \sum_{i\neq j} \delta p_{ti}\delta p_{tj}\rangle &=& \int \delta p_{t1}\delta p_{t2} \langle G_{12}\rangle d\omega_1d\omega_2\nonumber\\
 &+& \int \delta p_{t1}\delta p_{t2} \langle \langle f_1\rangle_n\langle f_2\rangle_n\rangle  d\omega_1d\omega_2.
\end{eqnarray}
To obtain (\ref{eq:dptEvolution}),  we use (\ref{eq:ptFluctExp}) and identify 
\begin{eqnarray}\label{eq:dptEquilib}
  \langle \delta p_{t1}\delta p_{t2}\rangle_e \equiv \!\! \int\!\!  \langle G_{12}^e\rangle\frac{\delta p_{t1}\delta p_{t2} }{\langle N(N-1)\rangle}  d\omega_1d\omega_2\quad\quad\quad \nonumber\\
  +  \!\! \int\!\! \delta p_{t1}\delta p_{t2} \frac{\langle\langle f_1\rangle_n\langle f_2\rangle_n\rangle}{\langle N(N-1)\rangle}  d\omega_1d\omega_2.
 %
%
\end{eqnarray}
Note that this derivation of (\ref{eq:dptEvolution}) follows a very similar derivation in Ref.\ \cite{Gavin:2016hmv}, section V., replacing $p_t$ there with $\delta p_t$ here. 

We understand the different  terms in these equations as representing distinct physical contributions. The first terms on the right sides of (\ref{eq:unres1}) and (\ref{eq:dptEquilib}) include all fluctuations {\em within} each event -- initial-state and dynamic. The second terms in these equations give the contribution to  $\langle \delta p_{t1}\delta p_{t2}\rangle$ from the variation of the average local equilibrium distribution from event to event. The first term in (\ref{eq:dptEquilib}) is likely small, and would vanish if the temperature and the transverse velocity were completely uniform on the freeze out surface. We expect variation from event to event to dominate  $\langle\delta p_{t1}\delta p_{t2}\rangle_e$.   

We now combine the solution (\ref{eq:pathIntegral2}) for $G_{12}$ with (\ref{eq:dptEvolution}) to compute $\langle \delta p_{t1}\delta p_{t2}\rangle$. The integral in (\ref{eq:dptEvolution}) yields 
\begin{eqnarray}\label{eq:dptArg1}
\int \delta p_{t1}\delta p_{t2} (G_{12}- G_{12}^e)d\omega_1d\omega_2  =  aS^2 +  bS,
 \end{eqnarray}
where $S$ is given by (\ref{eq:survival}), 
\begin{eqnarray}\label{eq:dptArg3}
a = \int \delta p_{t1}\delta p_{t2} \Delta G_{12}^0 d\omega_1d\omega_2,
 \end{eqnarray}
and 
\begin{eqnarray}\label{eq:dptArg2}
b = \int \delta p_{t1}\delta p_{t2} (X_{12}^0 + X_{21}^0) d\omega_1d\omega_2.
 \end{eqnarray}
We expect the first term on the right side of (\ref{eq:dptArg1}) to be the dominant contribution. In each event, the local equilibrium mean $p_t$ corresponding to $f^e$ is determined primarily by the parameter $T$, with small ``blue-shift'' corrections due to the radial component of $\mathbf{v}$. The variation of these parameters at the freeze out surface is likely small. We therefore approximate $\int dp_2\, p_{t2} X_{12}^0 = \int dp_2\, p_{t2}(\langle\delta f_1f^e_2\rangle_n - \langle \delta f_1\rangle_n f^e_2)\approx \langle p_t\rangle \int dp_2\, X_{12}^0$.  It follows that $b \approx 0$  because (\ref{eq:dptArg1}) depends on $\delta p_{t2} = p_{t2} -\langle p_t\rangle$.  Nevertheless, we stress that this argument only holds for $\langle \delta p_{t1}\delta p_{t2}\rangle$; the quantity $X_{12}^0$ need not generally be small.

To illustrate how $p_t$ fluctuations can be used to study thermalization, we simplify these integrals with a number of simplifying assumptions. We assume that freeze-out and particle formation occur at constant proper time as defined by the characteristic paths that satisfy (\ref{eq:path}). We focus on the contribution of long range correlations to $p_t$ fluctuations. Correspondingly, we take the underlying dissipation- and noise-free transverse flow corresponding to the local equilibrium distribution $f^e$ to be boost invariant along the beam direction.  This is appropriate in view of the observed rapidity independence of long range correlation features such as the ridge. In a fixed rapidity interval, these assumptions imply that the spatial integrals in $a$, $b$, and $\langle N(N-1)\rangle \approx \langle N\rangle^2$ all vary as $\tau^2$, so that the ratios in (\ref{eq:dptEvolution}) are roughly $\tau$ independent.  

We combine (\ref{eq:dptEvolution}) and (\ref{eq:dptArg1}) to estimate
\begin{eqnarray}\label{eq:dptFinal}
  \langle \delta p_{t1}\delta p_{t2}\rangle = \langle \delta p_{t1}\delta p_{t2}\rangle_o S^2 + \langle \delta p_{t1}\delta p_{t2}\rangle_e(1-S^2),\,\, 
\end{eqnarray}
where $S$ is the survival probability (\ref{eq:survival}).  Fluctuations start from an initial value $ \langle \delta p_{t1}\delta p_{t2}\rangle_o$ at the formation time $\tau_0$ and evolve toward the equilibrium value $\langle \delta p_{t1}\delta p_{t2}\rangle_e$. Rather than compute these quantities from (\ref{eq:dptEquilib}) and (\ref{eq:dptArg3}), we will estimate them as follows. 
The local equilibrium value $\langle \delta p_{t1}\delta p_{t2}\rangle_e$ is determined by fluctuations from event to event of the initial participant geometry. We estimate these fluctuations using the blast wave model from Ref.\ \cite{Gavin:2011gr}. This model provides excellent phenomenological agreement with a wide range of fluctuation, correlation and flow harmonic measurements at soft and hard scales \cite{Gavin:2008ev,Moschelli:2009tg,Gavin:2012if}.

The initial $p_t$ fluctuations are generated by the particle production mechanism, which we take to be string fragmentation for concreteness. Specifically, we approximate the early collision as a superposition of independent string-fragmentation ``sources.''  Each source contributes both  $p_t$ and multiplicity fluctuations, the latter characterized by 
\begin{equation}\label{eq:R}
{\cal R} 
%
= {{\langle N(N-1)\rangle - \langle N\rangle^2}\over{\langle N\rangle^2}};
\end{equation}
see \cite{Pruneau:2002yf}.  We expect both $\langle \delta p_{t1}\delta p_{t2}\rangle$ and ${\cal{R}}$ to vary inversely with the number of sources \cite{Gavin:2011gr}. 
%
Therefore, we write
\begin{equation}\label{eq:wnm}
    \langle \delta p_{t1}\delta p_{t2}\rangle_o \approx
   \mu
    {{\cal{R}}\over{1+\cal{R}}},
\end{equation}
%
where the $(1 + {\cal{R}})^{-1}$ factor accounts for the normalization of (\ref{eq:ptFluctExp}) to $\langle N(N-1)\rangle \equiv \langle N\rangle^2(1 + \cal{R})$ rather than $\langle N\rangle^2$. We fix the coefficient $\mu$ at each beam energy using PYTHIA by computing  $\langle \delta p_{t1}\delta p_{t2}\rangle$ and $\cal R$ for proton collisions.  We take ${\cal R}\propto (dN/dy)^{-1}$, and fix the proportionality constants to be consistent with the blast wave calculation. This ensures that $\langle \delta p_{t1}\delta p_{t2}\rangle_{e}$ and $\langle \delta p_{t1}\delta p_{t2}\rangle_{o}$ describe events with the same numbers of particles. 
%
%
%
\begin{figure}[t]
\includegraphics[width = \linewidth]{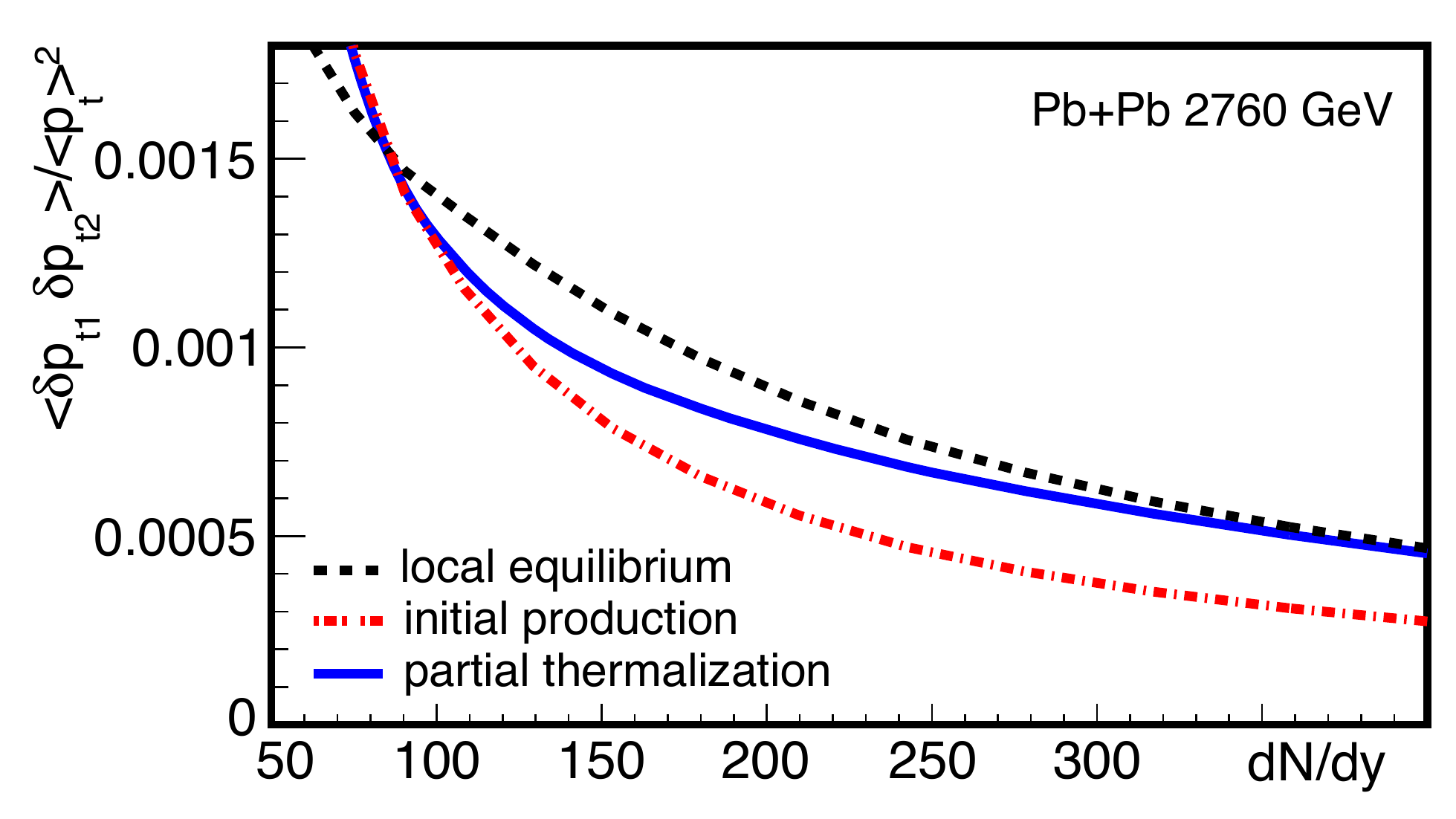}%
\caption{(color online) Pb+Pb fluctuations as a function of the charged-particle $dN/dy$ in the peripheral region where partial thermalization (solid curve) drives systems of increasing lifetime from the initial state (dash-dotted curve) to local equilibrium flow (dashed curves). 
\label{fig:Closeup}}
\end{figure}

We now ask whether the data in Fig.\ \ref{fig:Heavy} show signs of partial thermalization. As a benchmark, we compare calculations using our blast wave model of the event-wise fluctuations of thermalized flow. The dashed curves in Fig.\ \ref{fig:Heavy} show that blast wave results agree well over two orders of magnitude in beam energy for most of the centrality range, continuing the trend noted in Ref.\ \cite{Gavin:2011gr}. Nevertheless, our comparison here reveals a significant systematic deviation from the data in the most peripheral collisions. These events correspond to collisions with fewer than $\sim 50$ participants, compared to the maximum of $\sim 400$ in central collisions. This is precisely the sort of deviation one expects if the thermalization in these diffuse systems is incomplete \cite{Gavin:2003cb}.  

To estimate the extent of thermalization in peripheral heavy ion collisions, we compute the initial value $\langle \delta p_{t1}\delta p_{t2}\rangle_o$ using (\ref{eq:wnm}) and use our blast wave model to determine the equilibrium value $\langle \delta p_{t1}\delta p_{t2}\rangle_e$. Our blast wave $\langle p_t\rangle$ agrees with measured values at each energy within experimental uncertainties, so that partial thermalization does not appreciably alter $\langle p_t\rangle$ as in \cite{Gavin:2008ev}. We then use (\ref{eq:dptFinal}) to extract $S$ as a function of $dN/dy$, neglecting any possible beam energy dependence. The resulting solid curves in Fig.\ \ref{fig:Heavy} agree quite well given the simplicity of the model. While much more work is needed to draw quantitative conclusions, this agreement lends strong support to the possibility that these data are indeed measuring thermalization.

To clarify the effect of partial thermalization described by (\ref{eq:dptFinal}), we focus in Fig.\ \ref{fig:Closeup} on the peripheral region in $dN/dy$ in Pb+Pb collisions where the extracted $S$ drops from one to zero. Events producing the lowest $dN/dy$ have fluctuations closest to the initial distribution (\ref{eq:wnm}), shown as the dash-dotted curve.  We expect higher $dN/dy$ events to produce a larger collision volume that is more dense and longer lived. Consequently, the probability that a particle survives the collision without scattering $S$ should be smaller.  The values of $S$ we extract in Fig.\ \ref{fig:Heavy} agree with this expectation. Fig.\ \ref{fig:Closeup} shows that fluctuations computed using (\ref{eq:dptFinal}) approach  locally-thermal behavior $\langle \delta p_{t1}\delta p_{t2}\rangle_e$ above $dN/dy\approx 400$.  
%
%
\begin{figure}[t]
\includegraphics[width = \linewidth]{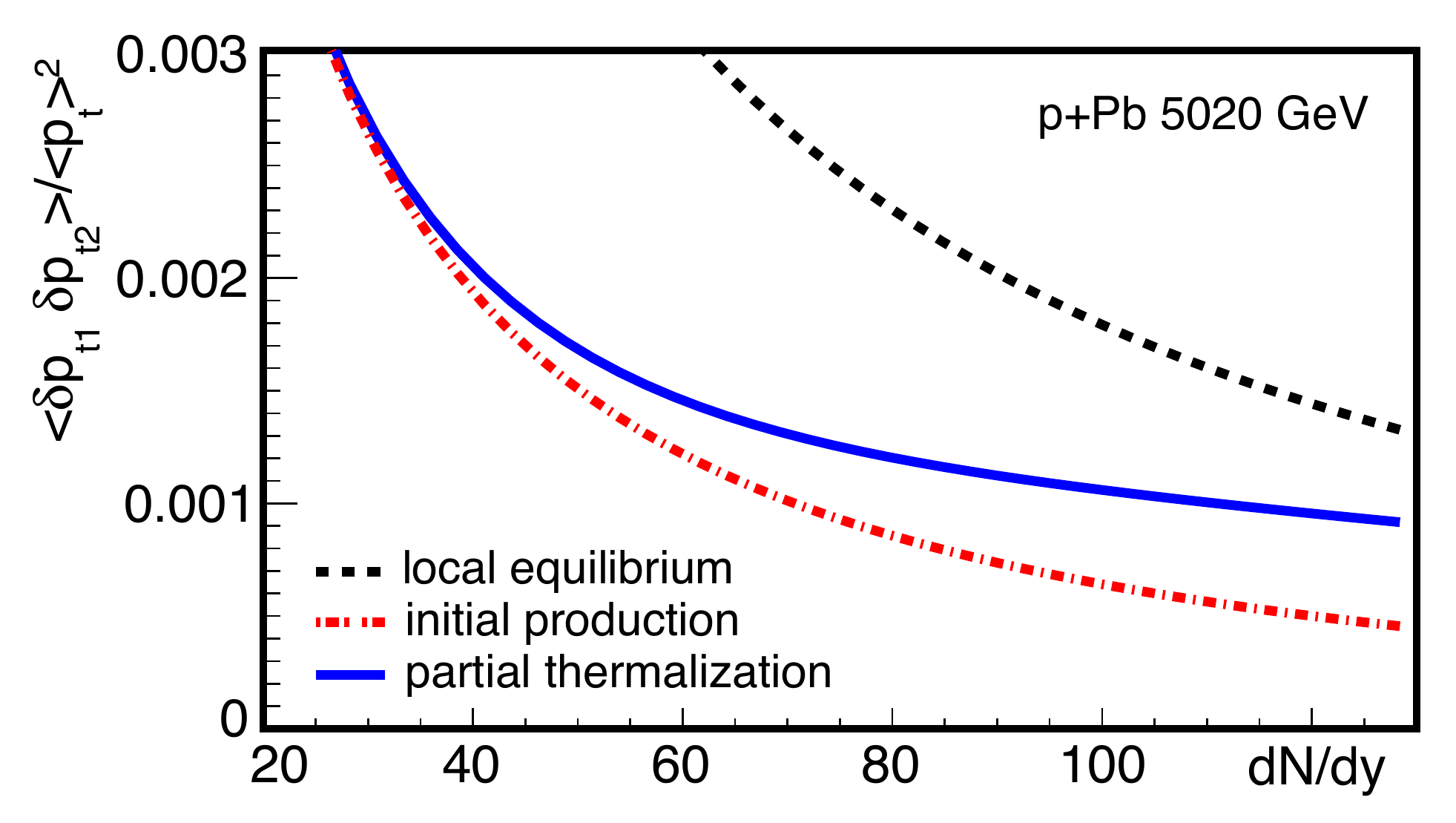}%
\caption{(color online) In p+Pb collisions partial thermalization becomes more prominent with higher multiplicity $dN/dy$. Extrapolated fluctuations for partial thermalization (solid curves) are compared to the initial particle production (dash-dotted curves) and local equilibrium flow (dashed curves). 
\label{fig:Light}}
\end{figure}

We comment that practicality drives our use of the blast wave model \cite{Gavin:2011gr}. We would prefer to compute $\langle \delta p_{t1}\delta p_{t2}\rangle_e$ using dissipation-free hydrodynamics with initial-state fluctuations. One could eventually combine three-dimensional hydrodynamics with, e.g, fluctuating IP-Glasma initial conditions. However, one must bear in mind that the statistics must be adequate to distinguish the solid and dashed curves at a range of energies as in Fig.\ \ref{fig:Heavy}. While a step in the right direction, Ref.\ \cite{Bozek:2012fw} has insufficient statistics to address questions posed here. Our experience suggests that this would take millions of events per beam energy.  


We argue that pA collisions are the best systems to look for partial thermalization.  To get a feel for the possibilities, we extrapolate our heavy ion estimate to a pA collision using the appropriate initial value from (\ref{eq:wnm}) and a blast wave calculation with parameters fit to pA data \cite{Preghenella:2013vna}. The result is shown in Fig.\ \ref{fig:Light}. Our partial thermalization result is obtained with the same $S$, but in the appropriate $dN/dy$ range. This extrapolation overlooks the fact that the dynamic evolution that determines $S$ in a pA collision is likely very different than that in the larger, longer lived and more dense Pb+Pb system.    

Is there anything we can say about the scattering processes that determine the survival probability $S$? A rigorous answer requires a detailed description of elastic and inelastic scattering that is beyond the scope of our exploratory work. However, we can obtain a rough estimate of the overall equilibration time scale $\nu^{-1}$ as follows.  Kinetic theory implies $\nu^{-1} \sim n\sigma v_{\rm rel}$, where the scattering cross section is $\sigma$, and $v_{\rm rel}$ is the relative velocity. If we take $\nu$ to be constant then the survival probability (\ref{eq:survival}) is $S \approx \exp[-\nu (\tau_F-\tau_0)]$. We estimate $S \approx 0.00435$ for the most central Pb+Pb collisions at 2.76 TeV. For a formation time $\tau_0 =0.6$~fm and a freeze out time $\tau_F=10$~fm, we find $\nu^{-1}\sim 1.7$~fm for the most central Pb+Pb collisions.  More realistically, if we take the density $n \propto \tau^{-1}$ to account for longitudinal expansion, but assume $\sigma v_{rel}$ to be constant, then $S \approx (\tau_0/\tau_F)^\alpha$,  where $\alpha = \nu_0\tau_0$ and $\nu_0$ is the initial value of $\nu(\tau)$. We then estimate the initial value $\nu_0^{-1}\sim 0.31$~fm, with a ten-fold increase as the system evolves. These values are consistent with the rapid thermalization required, e.g., by hydrodynamic analyses of flow harmonics.

\section{\label{sec:conclusion}Conclusion}
The primary aim in this paper is to develop theoretical and phenomenological tools for studying nonequilibrium aspects of correlation measurements. Our work is based on the Boltzmann-Langevin equation in the relaxation time approximation.  Our main result in Sec.\ \ref{sec:Roadmap} is the evolution equation for the two-particle phase space correlation function, Eq.\ (\ref{eq:G12}).

In Sec.\ \ref{sec:Beq} we discuss aspects of the Boltzmann equation and the relaxation time approximation -- linearized and not -- necessary for our work. The Boltzmann equation determines the relaxation of the one body phase space distribution $f$ to the local equilibrium distribution $f^e$.  Generally, $f^e$ is determined by nonlinear conditions (\ref{eq:RTAConstraint}) that impose the microscopic conservation laws for energy, momentum and any conserved quantum numbers. Linearizing reduces these conditions to  the requirement that the parameters $T$, $v$ and $\mu$ in $f^e$ satisfy effective ideal hydrodynamic equations. The difference between nonlinear relaxation time evolution and its linearized proxy are then evident by comparing the respective solutions (\ref{eq:pathIntegral}) and (\ref{eq:pathIntegralLinear}).

Importantly, the standard Boltzmann equation describes only dissipative processes that relax the system into an isotropic state devoid of correlations. Dynamic fluctuations preserve correlations by opposing dissipation and maintaining inhomogeneity;  see the discussion around Eq.\ (\ref{eq:CcorrF2}).  Even the equilibrium state must include fluctuations that balance dissipation.

We introduce Langevin fluctuations to the Boltzmann equation in Sec.\ \ref{sec:Roadmap} in order to describe nonequilibrium correlations. We derive the two body equation (\ref{eq:G12}) using methods for working with stochastic differential equations developed in Ref.\ \cite{Gavin:2016hmv} in the context of viscous hydrodynamics; see also \cite{van2011stochastic,gardiner2004handbook}. We introduce projection operators derived from the linearized Boltzmann equation in order to enforce the microscopic conservation laws. The resulting equation (\ref{eq:G12}) and its formal solution are capable of describing small fluctuations of a nonlinear average flow described by (\ref{eq:pathIntegral}). 

In deriving (\ref{eq:G12}) -- and solving it in Sec.\ \ref{sec:Collisions} -- we used the method of characteristics.  In Sec.\ \ref{sec:Beq} we used this method to find formal solutions (\ref{eq:pathIntegral}) and (\ref{eq:pathIntegralLinear}) for the one-body phase space distribution in terms of the survival probability $S$, (\ref{eq:survival}). We obtain a new formal solution for the two-body correlation function (\ref{eq:pathIntegral2}) that also depends on $S$. The characteristic method is the basis of familiar solutions to the one-body equation for Bjorken and Gubser flow, see e.g.\ Refs.\ \cite{Baym:1984np,Nopoush:2015yga}. However, we stress that  (\ref{eq:pathIntegral2}) applies much more generally. A characteristics-based approach can be implemented numerically for arbitrary initial conditions.  

To demonstrate the promise of these methods along with the practical issues involved, we use (\ref{eq:pathIntegral2}) to compute  the long range contribution to $p_t$ fluctuations in Sec.\ \ref{sec:Observables}. In Ref.\ \cite{Gavin:2003cb} we suggested that these fluctuations could be used to study thermalization. An appropriate body of data is now available \cite{Adams:2005ka,Abelev:2014ckr,Novak2013,Novak:2013zz}.  It is accepted that central collisions of large nuclei exhibit hydrodynamic flow. We therefore expect the first traces of thermalization to emerge in peripheral collisions, becoming more significant with increasing centrality  as the system lifetime increases.  
Indeed, peripheral collisions in Fig.\ \ref{fig:Heavy} show a systematic discrepancy with local equilibrium flow. Our partial thermalization model is in excellent accord with data over a range of energies.  

We hope our analytic result (\ref{eq:dptFinal}) for $\langle \delta p_{t1} \delta p_{t2}\rangle$ will motivate detailed cascade and hydrodynamic simulations. Our calculations in Fig. 1 show the subtle effect of thermalization on this observable in Pb+Pb collisions. One can use our result to judge the level of statistical accuracy needed for meaningful numerical simulations. 

As an aside, we comment that our evolution equation can be used to test simulation codes in regions where the answers are expected to overlap.  Our solution (\ref{eq:pathIntegral2}) can be compared to fluctuating hydrodynamic descriptions in the low density regime where both hydrodynamics and the Boltzmann equation generally give the same answers  \cite{Calzetta:1997aj,Kapusta:2011gt,Kumar:2013twa,Young:2014pka,Yan:2015lfa,Nagai:2016wyx, Gavin:2016hmv,Akamatsu:2016llw}.  Indeed, this was an important motivation for our work. It is also possible to use our approach to test cascade codes, provided that an equivalent description of scattering processes can be implemented. Baym's work on the one-body Boltzmann equation was used to obtain a simple analytic equation for the time evolution of the energy density \cite{Baym:1984np}.  That result was adapted in Ref.\ \cite{Gyulassy:1997ib} to test the authors'  parton cascade code. Our result  -- the first analogous analytic calculation of a  two-body quantity -- can be used to test cascade simulations of fluctuation quantities. 

Partial thermalization moves to center stage in pA collisions, as extrapolation of our heavy ion results suggests. More quantitative conclusions await further theoretical refinement and pA measurements. Meanwhile, we surmise that further information on the thermalization process is contained in many other observables on the periphery where hydrodynamics breaks down.  We aim to use the tools developed here to bring more of this information to light.   

{\em Note added:} Since the completion of this work two papers have argued for the measurement of $\langle \delta p_{t1} \delta p_{t2}\rangle$ in pA collisions citing contributions of different physical effects \cite{Bozek:2017elk,Osada:2017oxe}.  The recent transport model in Ref.\ \cite{Alqahtani:2017jwl} shows how we might incorporate mean field and other realistic effects into our work.


%

\begin{acknowledgments}
We thank Mauricio Martinez, Claude Pruneau, and Clint Young for useful discussions. We thank Bill Llope, John Novak, and Gary Westfall for discussions of preliminary STAR data from the RHIC Beam Energy Scan. This work was supported in part by the U.S. NSF grant PHY-1207687. 
\end{acknowledgments}

\bibliography{ptDiffusion_References}

\end{document}